# *The Determinants of Foreign Direct Investment (FDI)*
# *A Panel Data Analysis for the Emerging Asian Economies*

ATM Omor Faruq[1]

**Abstract**

In this paper, we explore the economic, institutional, and political/governmental factors in attracting Foreign Direct Investment (FDI) inflows in the emerging twenty-four Asian economies. To examine the significant determinants of FDI, the study uses panel data for a period of seventeen years (2002-2018). The panel methodology enables us to deal with endogeneity and other issues. Multiple regression models are done for empirical evidence. The study focuses on a holistic approach and considers different variables under three broad areas: economic, institutional, and political aspects. The variables include Market Size, Trade Openness, Inflation, Natural Resource, Lending Rate, Capital Formation as economic factors and Business Regulatory Environment and Business Disclosure Index as institutional factors and Political Stability, Government Effectiveness, and Rule of Law as political factors. The empirical findings show most of the economic factors significantly affect FDI inflows whereas Business Disclosure is the only important institutional variable. Moreover, political stability has a significant positive impact in attracting foreign capital flow though the impact of government effectiveness is found insignificant. Overall, the economic factors prevail strongly compared to institutional and political factors.

**JEL Classification: F21, F43,**

## Introduction

International trade and finance take a significant role in global economic system. One of the key reasons is globalization which is shaping the trade and finance in dimension. Cross border flow of capital, technology transfer, labor force distribution etc. provides a great degree of mutual benefits amongst the countries of different region. For example, FDI has assumed driving force behind economic growth in developing and emerging country. In recent years, the developing countries progressively formulate trade policy with intension to integrate with world markets.

The emerging Asian country have been experiencing increased flow of investment from other part of the world foreign direct investment inflows over the past decade. The emerging Asian market refers to those Asian economies which are experiencing considerable economic growth along with some characteristic of developed economy. High growth potential, market volatility, growing per capita, lower political stability, high risk are some common characteristics of the Emerging Asian market. The emerging and developing countries in Asia include mainly Bangladesh, Bhutan, Brunei Darussalam, Cambodia, China, Fiji, India, Indonesia, Kiribati, Lao P.D.R., Malaysia, Maldives, Marshall Islands, Micronesia, Mongolia, Myanmar, Nepal, Palau, Papua New Guinea,

---
[1] Department of Economics, Western Illinois University

Philippines, Samoa, Solomon Islands, Sri Lanka, Thailand, Timor-Leste, Tonga, Tuvalu, Vanuatu, and Vietnam etc. This country has been benefited from the FDI inflow in this region.

Lot of early research done to examine the determinants of FDI in different region of the world. However, there is still research gap to find out the ever-increasing foreign capital transferred to the developing market of Asia. Most of the early study heavily focus on economic factors only whereas the momentum of institutional quality and political stability can not be denied. Moreover, the impact of Global Financial Crisis on FDI inflow in these economies hardly examined in recent papers.

This paper examines the economic, institutional, and political/governmental factors in attracting Foreign Direct Investment (FDI) inflows in the emerging twenty-four Asian economies for the period of 2002-2018. To examine the significant determinants of FDI, the study uses panel data and random effect method.

## Literature Review

Many empirical pieces of literature focus on the factors affecting Foreign Capital Flow in different locations of the world. Those researchers attempted to find out the determinants of FDI from economic, institutional, and political perspectives. In this section, we discuss the theory related to FDI, FDI, and economic growth, etc. We also review early literature which investigates the factors behind FDI inflows in host countries.

### *Theory of FDI*

The concept of FDI is rooted in classical theories of international trade. In 1933, the Heckscher–Ohlin model (H–O model) was developed by Eli Heckscher and Bertil Ohlin at the Stockholm School of Economics. It is a general equilibrium mathematical model of international trade which is built on David Ricardo's theory of advantage. This theory suggests that a countries export is related to their cheap and available factors of production and import related to the countries' scarce resources (Kurtishi-Kastrati, 2013).

The theory of the Production Cycle developed by Vermon is related to the idea of foreign trade. The theory suggests four states of production: Innovation, growth, maturity, and decline. According to the theory, in the first stage, innovative products are made available for local consumption and then exported the surplus into the foreign market (Mullor-Sebastian, 1983).

The Internalization Theory explains how transitional companies grow and the motivation behind achieving foreign direct investment. The theory was first developed by Buckley and Casson, in 1976 which was further developed by the two Canadian economists, Stephen Hymer and John McManus. According to the theory, a firm proceed to internalization when perceived benefit overweight the cost. Firms that lead to foreign investment may incur diversified risk such as political and commercial due to the novelty of the foreign environment. These risks are known as costs of doing business abroad. (Buckley & Casson, 2009)

One of the most popular theories related to FDI is the Eclectic paradigm developed by professor Dunning (Dunning, 2000). This is the blend of three different theories of foreign direct investment.

In the first place, some firms have an ownership advantage in the form of ownership privileged in the market with limited resources. Moreover, intellectual property such as patents and trademarks provide a comparative advantage. In the second place, the advantage includes economic, social, and political may come from location. These advantages help the firm to grow beyond the national boundary. The last idea is Internalization which argues that if the first two conditions meet, it is more profitable for the company to move cross-market transactions.

The above-discussed theory gives us a solid understanding of the major theory of FDI and its implication. Next, we provide a brief literature review that investigates factors of FDI in various economies. To match with our research objectives, we discuss the early literature from three points of view: economic factors, institutional factors, and political/governmental factors.

The existing research findings on the determinants of FDI inflows in emerging economies are mixed and inconsistent. The main reason for the inconclusive result is the continuous changing of dynamism of the region.

Investment decision regarding FDI greatly influenced by the political stability and risk of certain location (Dunning & Lundan, 2008). Political instability interrupts firm's business operations and profitability. We also examined several other early research which establish relationship FDI and institutional factors such as Business Regulatory Environment, Business Interest in host country, rule of law etc.

*The factor of FDI: Economics, Institutional and Political*

The market size, trade openness, labor cost, lending rate, inflation, capital formation, economic stability, natural resource, etc. are considered as important economic factors which play a key role in the cross-country capital transfer.

In most of the empirical studies, the market size has been identified as a significant factor of FDI inflow to host countries. According to T.Bhavan who studied the FDI determinants in South Asian countries, the market size or population of the country is an important factor for attracting FDI. (Thangamani et al., 2010). Though this study considers population as the proxy variable for market size, most of the study examined the real GDP as the proxy variable in place of population. (Saini & Singhania, 2018) investigate the potential determinants of foreign direct investment in developed and developing countries on panel data using static and dynamic modeling. The study finds the FDI, and GDP has a positive impact. The reason behind this that with higher per capita people have a higher level of purchasing power which eventually gives a higher return to the foreign country investment.

Trade openness is measured as the sum of a country's exports and imports as a share of that country's GDP. A higher level of trade openness suggests the country's positive attitude towards foreign trade. (Uz Zaman et al., 2018) analyze the FDI and trade openness relationship for India Iran and Pakistan (1982-2012). To analyze the panel data, Fixed Effect and Pooled OLS techniques are used in this study. According to the empirical findings, the higher openness of trade positively impacts the inflow of foreign investment.

To overcome the macroeconomic shocks or negative effects of changes in home countries, the multinational corporation tries to shift the investment to the host country. (Valli & Masih, 2014) examine whether the theoretical relationship between FDI and inflation in South Africa. The empirical study demonstrates that a degree of causality exists between stable inflation levels and improved FDI inflows.

In emerging or developing economies, the congenial investment climate has a significant impact on attracting foreign investors. For a positive investment climate, the host country's Gross Capital Formation is considered as a required factor. However, (Krkoska, 2001) finds the capital formation is positively correlated with foreign investment along with domestic debt and capital market.

Natural Resource provides absolute advantage to a country as abundance of it sometimes related with factors of production. According (Hayat, 2018) suggests that after certain threshold, increased level of natural resource lead to negative significant effect on FDI.

In recent years, the link between institutions and FDI flows has been emphasized. It is assumed that foreign investors also consider the institutional and political factors when choosing the host country to establish their entity. Therefore, the host country should establish strong institutions and stable and effective government system which give certain level of confidence to foreign investors. We review several literatures which focus on institutional and political factors of FDI.

## Empirical Model

*Conceptual Framework*

The conceptual framework suggests that a country's flow of foreign capital depends on economic factors as well as political and institutional factors. The impact of the economic factors is assumed to be higher compared to the political and institutional factors. The inner crisscross sign defines the relationship between the depend on and independent variables. Here, we assume variables from different segments holistically work as an important parameter for FDI inflow. Besides, some degree of endogeneity may arise between the FDI and other variables.

*Empirical Model*

Based on the literature review and discussion of the conceptual framework, we propose three estimation models as follows, where the variables of interest are expected have a significant impact on FDI.

*Economic factor Model:*

$$\log(fdi_{it}) = \alpha + \beta_1 \log(gdp_{it}) + \beta_3 tradopp_{it} + \beta_4 lenrat_{it} + \beta_5 capfor_{it} + \beta_6 natres_{it} + \varepsilon_{it} \ldots \ldots \ldots \ldots (i)$$

Where,

$\log(fdi_{it})$ is the log of Foreign Direct Investment in Current US\$ for country i at time t.

$\log(gdp_{it})$ is the log of Gross Domestic Product in Current US\$ for country i at time t.

$infl_{it}$ is the rate of inflation in percentage form for country i at time t.

$tradopp_{it}$ is the ratio of trade as a percentage of GDP for country i at time t.

$lenrat_{it}$ is the lending rate in percentage form for country i at time t.

$capfor_{it}$ is the gross capital formation for country i at time t.

$natres_{it}$ is the natural resource rent in the percentage of GDP for country i at time t.

$\varepsilon_{it}$ is the error term over the time t.

*Political and Institutional factor Model:*

$$\log(fdi_{it}) = \alpha + \beta_1 gov.eff_{it} + \beta_2 pol.stab_{it} + \beta_3 rulaw_{it} + \beta_4 reg.qua_{it} + \beta_5 bizdis_{it} + \beta_6 bizreg_{it} + \varepsilon_{it} \cdots\cdots (ii)$$

Where,

$gov.eff_{it}$ is government effectiveness estimation for the country i at time t.

$pol.stab_{it}$ is the political stability estimation score for the country i at time t.

$rulaw_{it}$ is the rule of law estimation score for the country i at time t.

$reg.qua_{it}$ is the regulatory quality estimation score for the country i at time t.

$bizdis_{it}$ is the is business closure rating for the country i at time t.

$bizreg_{it}$ is the rating of the regulatory environment for the county i at time t.

$\varepsilon_{it}$ is the error term over the time t.

**Mixed Model**

$$\log(fdi_{it}) = \alpha + \beta_1 \log(gdp_{it}) + \beta_2 tradopp_{it} + \beta_3 capfor_{it} + \beta_4 natres_{it} + \beta_5 pol.stab_{it} + \beta_6 bizdis_{it} + \beta_7 gfc_{it} + \varepsilon_{it} \cdots\cdots\cdots\cdots (iii)$$

This model is developed based on the finding of the other two models. We just included some variables from models 1 and 2. We also control the year 2008 and 2009 to measure the impact of the Global Financial Crisis on FDI inflow.

The study analyzed the data from the emerging Asian country. The panel data estimation is employed considering the dynamic behavior of the parameters. Again, the ordinary least square method may provide efficient α and β estimation. The main objectives behind using panel data are to incorporate maximum information and avoid any sample biased due to the time series nature relationship between the FDI and independent variables.

There are three different models for panel data method a) Common Constant (Pooled OLS) b) Fixed effects c) Random effect. In this study, we use the Random effect model over pooled OLS and fixed-effect model. The random assume that in terms of error term each country is unique whereas the fixed effect suggests that each country differs in terms of intercept term. We choose to use a random effect model based on the Hausman specification test.

## Description of Data and Descriptive Statistics

To study the effect of determinants on the FDI, we include yearly observation for seventeen years period (2002-2018) for twenty-four emerging Asian countries. Two databases: World Development Indicator (WDI) and World Governance Indicator (WGI) from World Bank are used for the required panel data. We use different indicators which can be categorized into three broad categories: Economic factor, Institutional factor, and Political factors. The variables description is as follows:

| Variable | Expected Sign | Description |
|---|---|---|
| **Foreign Direct Investment** | | Foreign direct investment, net inflows (BoP, current US$). |
| **Economic Factor** | | |
| **Market Size** | (+) | To measure Market size, we use GDP (BoP, current US$). |
| **Trade openness** | (+) | Trade (% of GDP) |
| **Total natural resources rents (% of GDP)** | (+) | The sum of oil rents, natural gas rents, coal rents (hard and soft), mineral rents, and forest rents. |
| **Lending Rate (%)** | (-) | The bank rate usually meets the short- and medium-term financing needs of the private sector. |
| **Inflation** | (-) | Inflation, consumer prices (annual %) |
| **Gross Capital Formation (%)** | | Indicates the savings of the country as % of GDP |
| **Institutional Factor** | | |
| **The business extent of disclosure index (Low 0-High 10)** | (+) | The disclosure index measures the extent to which investors are protected through the disclosure of ownership and financial information. |
| **CPIA business regulatory environment rating (1=low to 6=high)** | (+) | Assess the extent to which the legal, regulatory, and policy environments help or hinder private businesses in investing, creating jobs, and becoming more productive |
| **Governmental Variables** | | |
| **Political Stability (Low 2.5 to 2.5 High)** | (+) | Measure perceptions of the likelihood of political instability and/or politically motivated violence, including terrorism. |
| **Government Effectiveness (Low 2.5 to 2.5 High)** | (+) | Capture perceptions of the quality of public services, the quality of the civil service and the degree of its independence from political pressures, the quality of policy formulation and implementation, and the credibility of the government's commitment to such policies. |
| **Rule of Law (Low 2.5 to 2.5 High)** | (+) | captures perceptions on rules of society, quality of contract enforcement, property rights, the police, and the courts, as well as the likelihood of crime and violence. |

*Table 1: Descriptive Statistics of variables of the study*

| Variable | Observation | Mean | Min | Max |
|---|---|---|---|---|
| GDP | 408 | 804 Billion | 7.959e+08 | 1.087e+13 |
| Lending Rate | 304 | 9.83 | 0.99 | 35.51 |
| Inflation | 408 | 6.57 | -25.13 | 45.48 |
| Trade Openness | 408 | 92325 | 0.17 | 437.32 |
| Natural Resource | 408 | 9.89 | .00019 | 55.34 |
| Business Reg. Environment | 137 | 3.36 | 2.5 | 4.0 |
| Business Disclosure Index | 336 | 6.70 | 2.00 | 10 |
| Political Stability | 408 | -2.94 | -2.50 | 2.5 |
| Government Effectiveness | 408 | .11 | 0.00 | 1 |
| Rule of Law | 408 | -06 | -1.73 | 1.84 |

*Table 2: Descriptive Statistics of variables of the study*

|  | FDI | GDP | Infla. | Nat.Res | Tradopp | Rullaw | Pol.Stab | Gov.eff |
|---|---|---|---|---|---|---|---|---|
| FDI | 1.00 | | | | | | | |
| GDP | .0.71 | 1.00 | | | | | | |
| Infla. | -0.11 | -0.20 | 1.00 | | | | | |
| Nat.Re | -0.12 | -0.211 | 0.27 | 1.00 | | | | |
| Tradopp | 0.045 | -0.24 | -0.17 | 0.05 | 1.00 | | | |
| Rullaw | 0.05 | 0.21 | -0.29 | -0.05 | 0.34 | 1.00 | | |
| Pol.Stab | 0.02 | 0.09 | -0.13 | 0.25 | 0.40 | 0.70 | 1.00 | |
| Gov.eff | 0.15 | 0.25 | -0.31 | -0.31 | 0.41 | 0.94 | 0.69 | 1.00 |

*Empirical Results*

*Table: 3: Economic Factors Model*

| Variables | Log (FDI) |
|---|---|
| **Constant** | -13.405*** |
|  | (3.488) |
| **Market Size | Log (GDP)** | 1.276*** |
|  | (0.127) |
| **Trade Openness** | 0.011 |
|  | (0.009) |
| **Lending Rate** | -0.027 |
|  | (0.024) |
| **Gross Capital Formation** | 0.053*** |
|  | (0.009) |
| **Natural Resource** | 0.031*** |
|  | (0.012) |
| **No of Observation** | 290 |
| **$R^2$** | 0.463 |
| **Adjusted $R^2$** | 0.451 |
| **F-Statistics** | 210.930*** |
| **Note** | *p<0.1; **p<0.05; ***p<0.01 |

From table 3, we find that Market Size, Gross Capital Formation, Natural resources are the significant economic factors that attract Foreign Investment in the emerging Asian economies. The significance of trade openness and lending rate is not established in this model. The R-squared value is 0.463 which indicates the model explains 45.30% variation in FDI inflows in the emerging Asian economy.

Following equation (2), we examine what economic and institutional factors work as determinants of FDI. The result is reported in Table: 4.

*Table: 4: Economic and Institutional Factors Model*

| Variables | Log (FDI) |
|---|---|
| **Constant** | -19.398*** |
|  | (1.413) |
| **Government Effectiveness** | 0.984 |
|  | (0.642) |
| **Political Stability** | 0.600** |
|  | (0.240) |
| **Rule of Law** | -1.213* |
|  | (0.679) |
| **Regulatory Quality** | -0.731 |
|  | (0.528 |
| **Business Disclosure Index** | 0.088 |
|  | (0.093) |
| **Business Regulatory Environment** | 0.207 |
|  | (0.303) |
| **No of Observation** | 134 |
| **$R^2$** | 0.361 |
| **Adjusted $R^2$** | 0.331 |
| **F-Statistics** | 22.009*** |
| **Note** | *p<0.1; **p<0.05; ***p<0.01 |

From the empirical result, except Rule of law and Political Stability, no other variables seem significant in attracting FDI. Both are significant at a 10% significant level. However, the sign of the coefficient of rule of law is counter intuitive. It was expected the stronger rule of law prevails; the foreign investors would be more likely to transfer their capital from their home country to the host country. This model suggests that the institutional factors have no impact on the flow of FDI.

Following equation (iii), to find out all significant determinants of FDI, we control different economic, institutional, and political factors in model 3. The result is reported in Table: 5.

*Table: 5: Mixed Factors Model*

| Variables | Log (FDI) |
|---|---|
| **Constant** | -5.427** |
|  | (2.369) |
| **Market Size \| Log (GDP)** | 0.967*** |
|  | (0.092) |
| **Trade Openness** | 0.004* |
|  | (0.002) |
| **Gross Capital Formation** | 0.042*** |
|  | (0.008) |
| **Natural Resource** | 0.032*** |
|  | (0.008) |
| **Political Stability** | 0.238** |
|  | (0.117) |
| **Business Disclosure Index** | 0.042*** |
|  | (0.008) |
| **Global Financial Crisis (1)** | -0.092 |
|  | (0.111) |
| **No of Observation** | 325 |
| **R²** | 0.362 |
| **Adjusted R²** | 0.348 |
| **F-Statistics** | 191.083*** |
| **Note** | *p<0.1; **p<0.05; ***p<0.01 |

The result of the regression model shows that Market Size, Gross Capital Formation, Natural Resource, Business Disclosure Index has a positive significance to FDI at 10% significance level. Political Stability has a positive significance to FDI at a 5% significance level. Moreover, Trade Openness is significant at a 5% significant level. As we control the years 2008 and 2009 to measure the impact of the Global Financial Crisis on FDI, we find negative relation with FDI. However, the coefficient of the Global Financial Crisis is not significant.

The R-square value is 0.362 which indicates 36.2% variation in the data is explained by the model. The Adjusted R-squared values suggest that there are no additional factors that could affect FDI inflow in the emerging Asian economies.

**Discussion and Conclusion**

The study uses holistic approach to bring economic, institutional, and political determinants of FDI to better explain increased flow of FDI in the developing Asian economies. We use a seventeen-year data for twenty-four countries to overcome any sample biased. Traditional economic variables such as market size, trade openness, gross capital formation, natural resource plays a dominating role in attracting foreign investors. The main policy implications are that the growing GDP per capita seems profitable for the investors. Moreover, sound trade policy is required for attracting continuous flow of capital. Though the political factors such as government effectiveness, political stability, rule of law were expected to be significant, only the significant of political stability is established. On the other hand, Business Disclosure Index is significant positive impact on FDI which suggests that multinational corporation emphasizes more cautious about business regulation, gauge regulatory outcomes, legal protection of property, the flexibility of employment regulation, and the tax burden in the emerging Asian countries.

The world observed unprecedent economic shock-Global Financial Crisis-in 2008-09. It was one of the worst financial shocks after the Great Depression. Likewise rest of the world, most of the emerging country in Asia also faced low economic growth, downturn in FDI flow, high inflation, high rate of unemployment etc. However, the study does not find any negative significance of the GFC on FDI. Further study could be done to fine the reason.

The world investment scenario is changing constantly. According to (*UNCTAD*, World Investment Report, 2019), the Asia is renamed as the world's largest receiver of FDI gaining 39% of global inflows in 2018, which is 33% more than previous year. It was expected that the flow would be higher in coming years. In another study IMF estimated that the the total GDP of the Asian economy would exceed the rest of the world in 2020(*WEF, 2019*.). However, the world currently going through a COVID-19 pandemic. The world health system is about to collapsed with growing number of case and mortality. At the same time, the economic growth becomes slow down, unemployment increased, transfer of capital not promising. Therefore, the future study could include impact of pandemic, technological progress, and other important variables on the FDI inflow in the emerging Asian markets.

Moreover, the study has some limitation such as endogeneity between variable, missing values etc. which could be addressed in future research.

# Appendix

## Log (FDI) and Log (GDP): 2002 vs 2018

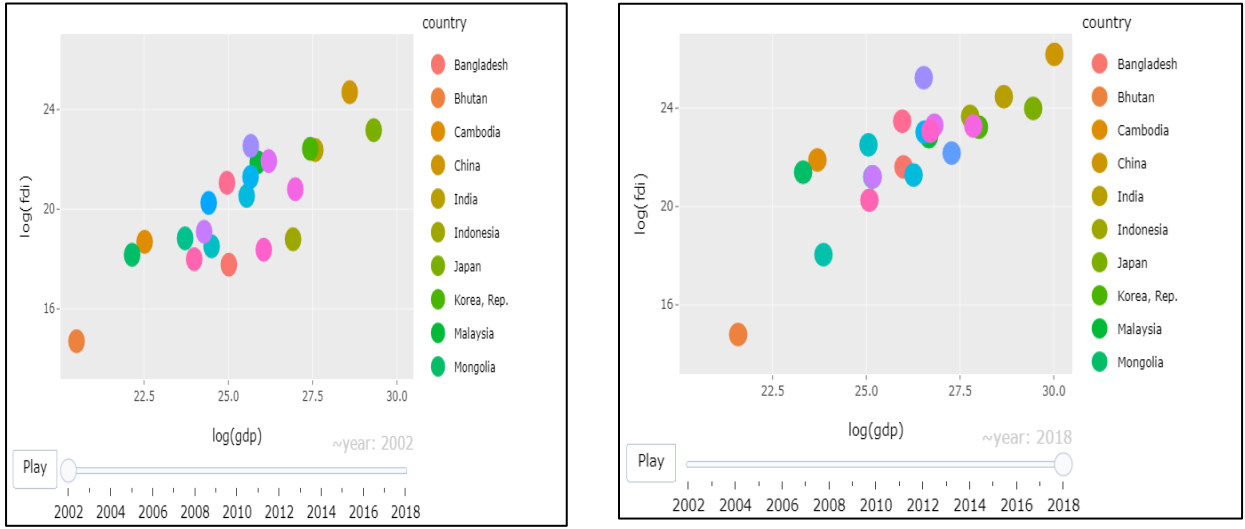

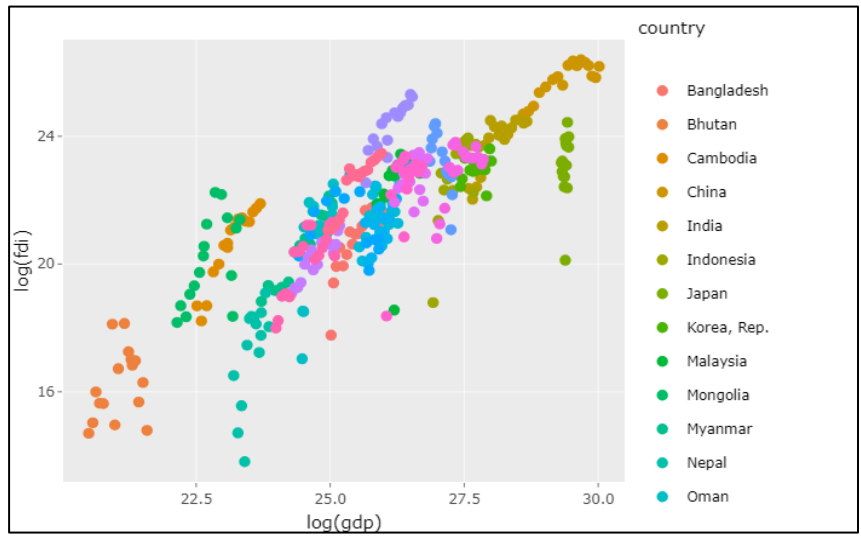

Market Size|GDP vs FDI inflow in the emerging Asian Country

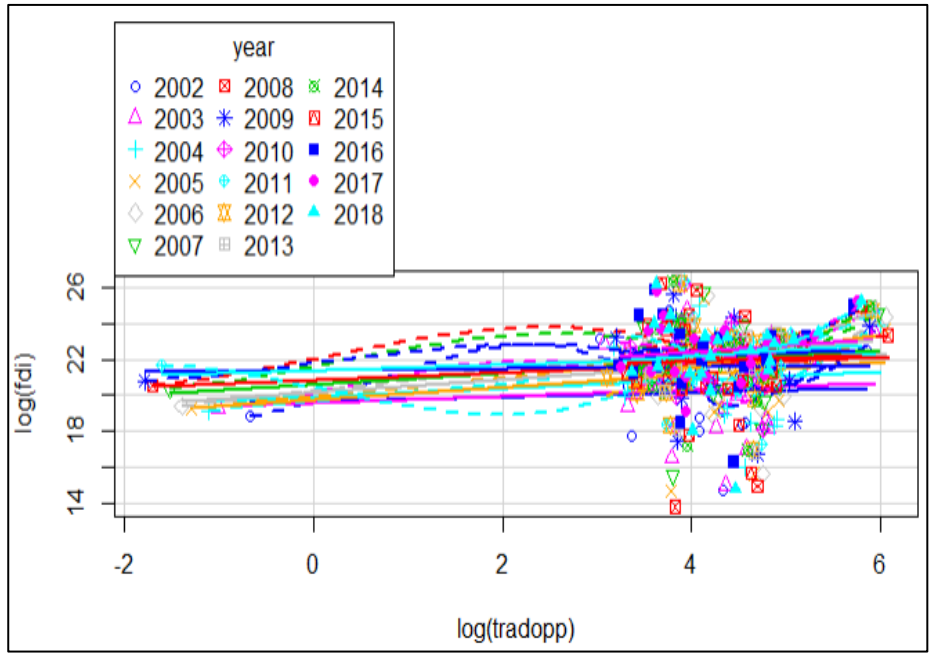

Trade Openness vs FDI: 17 Years Data

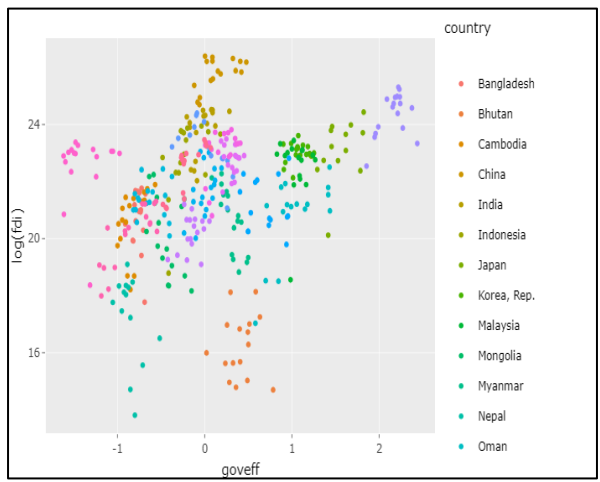
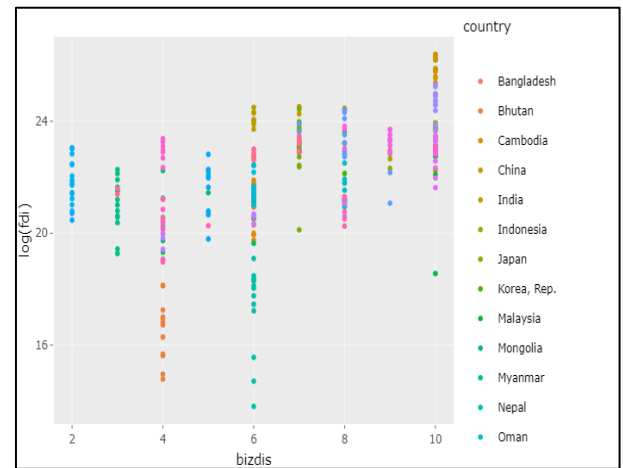

**Effective Government vs FDI inflow**
**Business Disclosure vs FDI inflow**